\pdfoutput=1
\documentclass[conference]{IEEEtran}
\ifdefined\pdfsuppresswarningpagegroup
\fi

\usepackage{amsmath}
\usepackage{booktabs}
\usepackage{cite}
\usepackage{graphicx}
\usepackage{tabularx}
\usepackage{url}
\usepackage{xcolor}

\newcommand{\ocudu}{OCUDU}
\newcommand{\us}{\ensuremath{\mu\mathrm{s}}}

\newcommand{\mode}[1]{\texttt{#1}}
\newcommand{\cfg}[1]{\texttt{\footnotesize #1}}

\usepackage[hidelinks,breaklinks]{hyperref}

\begin{document}

\title{GPU-Resident CUDA Acceleration for OCUDU 5G PHY and O-RAN Fronthaul: Architecture and Preliminary Performance}

\author{
\IEEEauthorblockN{Matthew Pennybacker}
\IEEEauthorblockA{DeepSig, Inc.}
\and
\IEEEauthorblockN{Wan Liu}
\IEEEauthorblockA{DeepSig, Inc.}
\and
\IEEEauthorblockN{Andriy Kharchenko}
\IEEEauthorblockA{DeepSig, Inc.}
\and
\IEEEauthorblockN{Timothy O'Shea}
\IEEEauthorblockA{DeepSig, Inc.}
}

\maketitle

\begin{abstract}
This paper describes DeepSig's CUDA-based acceleration backend for the \ocudu{} physical layer and O-RAN fronthaul path, integrated through acceleration interfaces that are largely independent of the underlying acceleration mechanism. The design accelerates PDSCH, PUSCH, SRS, PRACH, split-8 lower-PHY transforms, and O-FH IQ compression/decompression while preserving existing factories, resource-grid interfaces, PRACH-buffer interfaces, and channel processors. CUDA-visible grids, device-side softbit buffers, stream events, pinned staging buffers, and managed-memory policies keep data resident on the accelerator when the platform and radio split permit it. On an NVIDIA DGX Spark platform with a GB10 GPU and ARM CPU host, representative measurements with CPU baselines pinned to high-capacity cores show up to 10.3x PUSCH speedup, 2.7x PDSCH speedup, 19.7x split-8 low-PHY RX speedup with slot-shaped batching and scattered mapped zero-copy, 91.4x O-FH BFP12 decompression speedup, and 28.8x PRACH detector speedup against the production CPU path, with CPU and GPU 10\% BLER thresholds agreeing to within 0.064~dB in the tested PUSCH sweeps. The same resident pipeline provides an execution substrate for AI-RAN, allowing machine-learned channel estimation, neural receivers, and AI-native air-interface research to run beside standards-compliant baseband kernels.
\end{abstract}


\begin{figure*}[t]
\centering
\includegraphics[width=\textwidth]{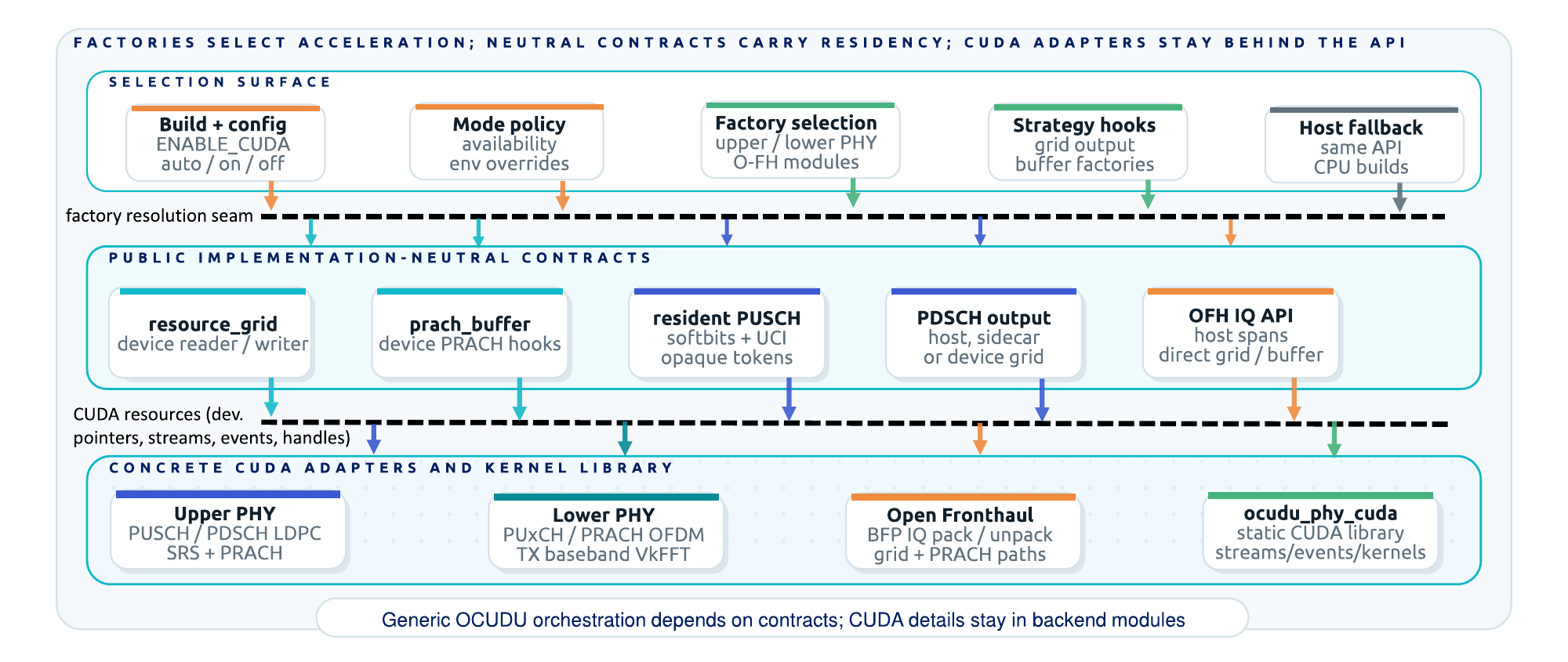}
\caption{Software encapsulation for CUDA acceleration. Existing factories select acceleration, neutral contracts carry device-resident handoffs, and CUDA allocation, synchronization, libraries, and kernels remain inside backend modules.}
\label{fig:softwaremap}
\end{figure*}

\section{Introduction}

GPU implementations of Viterbi and LDPC decoding appeared simultaneously and independently in 2008--09 across supercomputing, signal-processing, and communications venues. In particular, the use of hardware acceleration in a software-defined radio stack can be traced to pioneering work~\cite{falcao2008massive}. The follow-up work~\cite{falcao2009gpus} argued explicitly that LDPC decoding is normally served by VLSI delivering excellent throughput at significant cost, and proposes a flexible CUDA multi-codeword decoder instead, reporting throughputs above 100 Mb/s with BER curves comparable to ASIC solutions. Implementations of particular SDR-stack components were presented in~\cite{ji2009massively} (LDPC), \cite{wu2010turbo} (LTE), and \cite{wu2011mimo} (MIMO). Kim \emph{et al.}~\cite{kim2010sdrgpu} described a GPU-as-SDR-backend, proposing a hardware/software platform for an SDR modem on GPU in place of conventional DSPs and/or FPGAs. The integration work nearest to the \ocudu{} initiative, shifting from a standalone fast kernel to an accelerator inside a framework, was described in~\cite{horrein2011adapting,horrein2012integration,plishker2011applying}, and on the CPU side within Sora~\cite{tan2011sora}. Structural precedents for hardware-accelerated SDR with one buffer contract and host, CUDA, and HIP backends were described in~\cite{hitefield2016flowgraph,sorber2021accelerator}.

Open and programmable radio-access-network software is increasingly expected to support two concurrent workloads. First, it must satisfy conventional 5G NR real-time deadlines for PDSCH, PUSCH, PRACH, SRS, LDPC coding, and lower-PHY processing as specified by NR physical-layer procedures~\cite{ts38211,ts38212,ts38214}. Second, it must provide a practical execution environment for AI-RAN research and deployment, where machine-learned channel estimation, neural detection, channel-state analytics, and physical-layer AI functions execute inline with the modem~\cite{airan-alliance,ts38843}; the surrounding O-RAN architecture, interfaces, and open research problems are surveyed in~\cite{ai-ran}. An AI sidecar attached after CPU host-side baseband processing adds copies and scheduling jitter, and it denies neural components direct access to the grids, channel estimates, softbits, decoded codeblocks, and fronthaul IQ data they operate on.

The work described here, a CUDA~\cite{cuda} acceleration backend developed by DeepSig, adds GPU execution to the modem while retaining its existing architecture. \ocudu{} already instantiates its processing components through \emph{factories}: configuration-driven builder objects that construct a concrete implementation behind a stable interface at startup, so calling code never names the implementation it receives. Rather than introducing a new integration mechanism, this work extends those existing factories and optional interfaces to select the CUDA implementations, leaving high-level gNB logic untouched. CPU implementations remain available, and most knobs follow a common \mode{auto}, \mode{enabled}, and \mode{disabled} policy. The design supports split-8 SDR operation, split-7.2 O-FH operation following O-RAN user-plane compression formats~\cite{oran-wg4-cus}, and CPU fallback. Accelerated implementations also differ in how memory is shared between the CPU host and the GPU: unified-memory devices, where both processors access one coherent memory, can use CUDA-visible managed resource grids, while discrete GPUs, whose device memory is separate from host memory, use page-locked (pinned) host staging buffers and explicit H2D/D2H transfers where the radio or NIC boundary requires host memory. Section~\ref{sec:memory} develops these memory modes as accelerated implementation options.

\ocudu{} builds on the Software Radio Systems open-source RAN lineage, from srsLTE's LTE experimentation platform to the srsRAN Project 5G CU/DU codebase that SRS has positioned as the basis for \ocudu{}~\cite{srslte-wintech,srsran-project,srs-ocudu}. It also already contains a broader acceleration story: NEON, AVX2, and AVX-512 O-FH compression backends, plus DPDK bbdev-based FEC offload hooks for ACC100, ACC200, and VRB1 families. The proposed CUDA work is therefore one backend in an \ocudu{} TSC-defined hardware-acceleration architecture, with common benchmarks and interfaces so \ocudu{} can choose the best backend for each platform.

Data residency drives most of the design decisions. A GPU PUSCH path that copies LLRs to host after demodulation and copies them back for LDPC decode gives up much of the benefit of acceleration, so the implemented chain keeps intermediate PUSCH softbits on the device, fuses deinterleaving and rate dematching where possible, runs LDPC on the accelerator, and performs transport-block reassembly and CRC handling before the final payload result is synchronized. Similar ideas apply to PDSCH, low-PHY, SRS, PRACH, and O-FH compression.

\subsection{Relation to NVIDIA Aerial}
NVIDIA Aerial is an adjacent CUDA-based GPU-RAN option. The current Aerial documentation describes CUDA-accelerated RAN libraries for L1/L2, FAPI support for PHY-MAC communication, and cuPHY as an inline GPU accelerator for 5G gNB workloads~\cite{nvidia-aerial}. One viable accelerated \ocudu{} architecture is therefore to use an Aerial L1 implementation underneath an SCF FAPI adapter while retaining \ocudu{} L2/L3, CU/DU control logic, configuration, monitoring, and integration tooling above it.

The CUDA-accelerated backend design described here has a different purpose. It lives inside \ocudu{} as a modular acceleration backend rather than an external L1 replacement, aiming at preserving the unified \ocudu{} L1/L2/L3 codebase, keeping CPU and accelerated paths under the same factories and tests, and avoiding a requirement that every deployment adopt a single vendor stack, platform profile, or network-adapter topology. CUDA is the first implementation target evaluated here, but the factory and buffer abstractions are intended to support other hardware-acceleration backends as well. AI-RAN deployments, where AI functions may replace or augment pieces inside PHY, MAC, DU, or CU, especially benefit from the same flexibility.

\section{Software Architecture}
\subsection{Factory-Based Modular Acceleration}
Acceleration is configured through the same factories \ocudu{} already uses to assemble its CPU processing chains. Fig.~\ref{fig:softwaremap} organizes the design into three layers: a selection surface at the top, public implementation-neutral contracts in the middle, and concrete CUDA adapters at the bottom. The selection surface combines the CUDA build option, per-block acceleration-mode policy, and factory selection for the upper-PHY, lower-PHY, and O-FH modules, with host fallback available through the same APIs. All acceleration-mode settings accept \mode{auto}, \mode{enabled}, and \mode{disabled}; \mode{auto} selects an accelerated implementation when one is built and available---in the backend presented here, the DeepSig CUDA implementation---and otherwise falls back to the CPU path; the selection machinery itself is agnostic to which acceleration mechanism supplies the implementation. Algorithmic selection remains separate from acceleration enablement: for example, \cfg{ldpc\_decoder\_algorithm} selects \mode{auto}, \mode{boxplus}, or \mode{min\_sum} inside the accelerated decoder.

The middle layer of Fig.~\ref{fig:softwaremap} carries residency across component boundaries without naming CUDA: the resource-grid reader and writer interfaces expose optional device mapping hooks, the PRACH buffer exposes analogous device hooks, and resident PUSCH softbit and PDSCH output contracts hand off device-resident data as opaque tokens, yet high-level code can still treat the grid as a normal resource grid, so component boundaries are unchanged. Everything CUDA-specific stays in the bottom layer of the figure: allocation, streams, events, library handles, and kernel launches live in CUDA-specific implementation classes such as the CUDA-visible resource grid, accelerated lower-PHY modulators/demodulators, PUSCH/PDSCH processors, SRS and PRACH implementations, and O-FH compression adapters, all backed by a common static CUDA kernel library.

\subsection{End-to-End Dataflow}
\begin{figure*}[t]
\centering
\includegraphics[width=\textwidth]{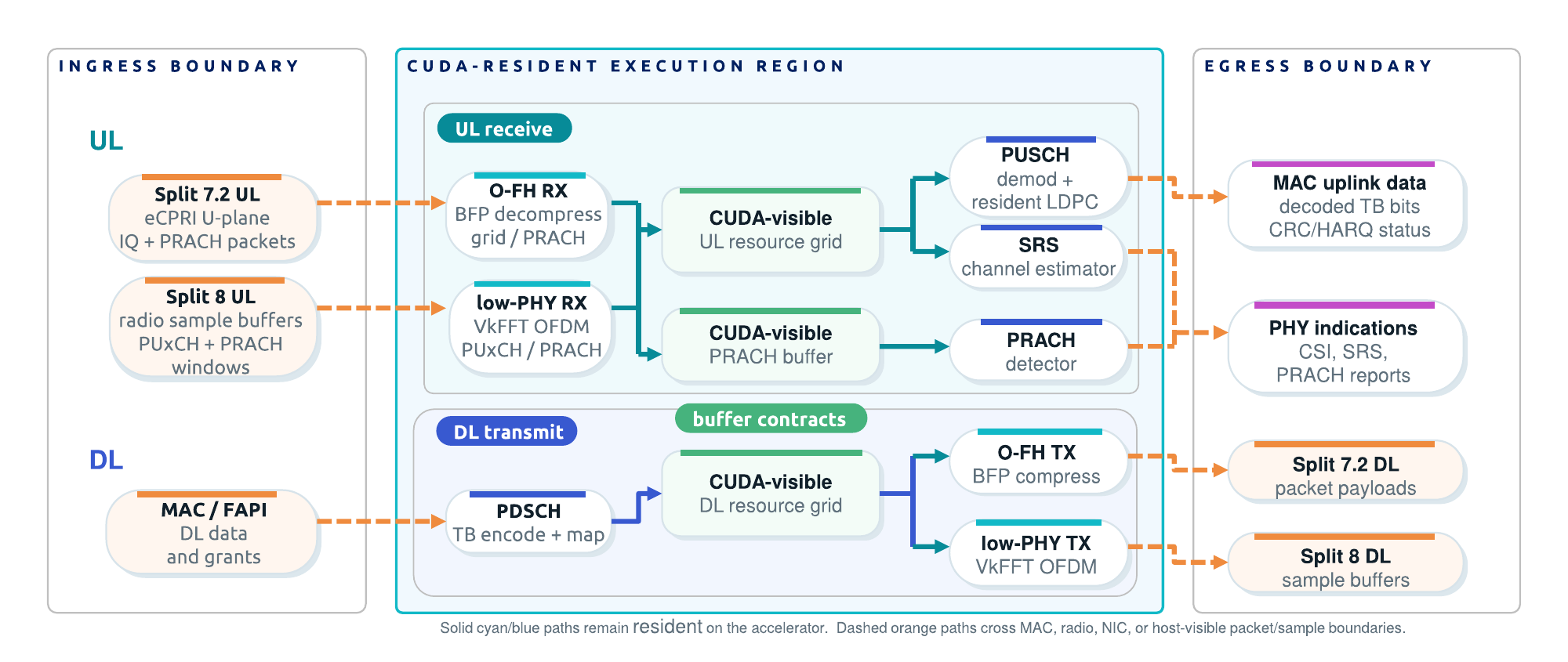}
\caption{CUDA-resident pipeline for split 7.2 and split 8. Uplink converges on CUDA-visible grids or PRACH buffers, PUSCH emits decoded transport-block bits and reports to MAC/scheduler, and PDSCH remains resident until the selected radio boundary.}
\label{fig:pipeline}
\end{figure*}

Fig.~\ref{fig:pipeline} summarizes the implemented residency model as three regions: an ingress boundary where packets, samples, and MAC/FAPI requests arrive in host memory; a CUDA-resident execution region where solid paths stay on the accelerator; and an egress boundary where decoded data, indications, and radio-bound buffers cross back to the host. Dashed paths in the figure mark those host crossings. In split 7.2 uplink, entering at the top-left of Fig.~\ref{fig:pipeline}, Ethernet and eCPRI/U-plane headers are parsed on the CPU host, but compressed IQ payloads can be decompressed by CUDA kernels directly into a CUDA-visible uplink resource grid. PUSCH and SRS can consume that grid through the reader's device pointer and ready event. PRACH can use an analogous CUDA-visible PRACH buffer. In split 8, the SDR boundary is host baseband samples; the accelerated low-PHY RX path performs OFDM demodulation and writes a CUDA-visible grid before upper-PHY consumers run.

On the downlink, along the bottom of Fig.~\ref{fig:pipeline}, MAC and FAPI inputs remain host-side control-plane data at the ingress boundary. PDSCH encoding, LDPC, rate matching, modulation, scrambling, precoding, and mapping execute on the GPU and populate a CUDA-visible downlink grid. Split-8 low-PHY TX can transform this grid into time-domain samples. Split-7.2 O-FH TX can compress the grid into packet payload buffers. Today, the NIC/radio host boundary still requires host-visible packet or SDR buffers, but the design minimizes the number and size of those host crossings.

The ingress and egress boundaries in Fig.~\ref{fig:pipeline} differ by split. The SDR split-8 path receives and transmits contiguous sample buffers through the radio plugin, so lower-PHY CUDA acceleration focuses on rapidly converting those buffers into or out of the resource-grid representation. The O-FH split-7.2 path receives packetized compressed IQ; header parsing and packet scheduling stay on the CPU today, but a decompressed host grid can be avoided. The compression interfaces expose both host-buffer and device-grid entry points so the same compressor can serve a packet payload path and a fully device-resident benchmark path without leaking CUDA into O-FH scheduling logic.

\section{Memory and Synchronization Model}\label{sec:memory}
The resource grid is the key abstraction used to avoid exposing CUDA to high-level logic. A CUDA-visible grid implements the normal host reader/writer API and adds optional device mapping functions: the reader can advertise device-readable CBF16 storage and a producer-ready event, while the writer can advertise device-mappable BF16 storage. Consumers call a preparation hook with their stream; the implementation orders the consumer stream against producer events using device-side synchronization rather than blocking the CPU.

\begin{figure*}[t]
\centering
\includegraphics[width=0.72\textwidth]{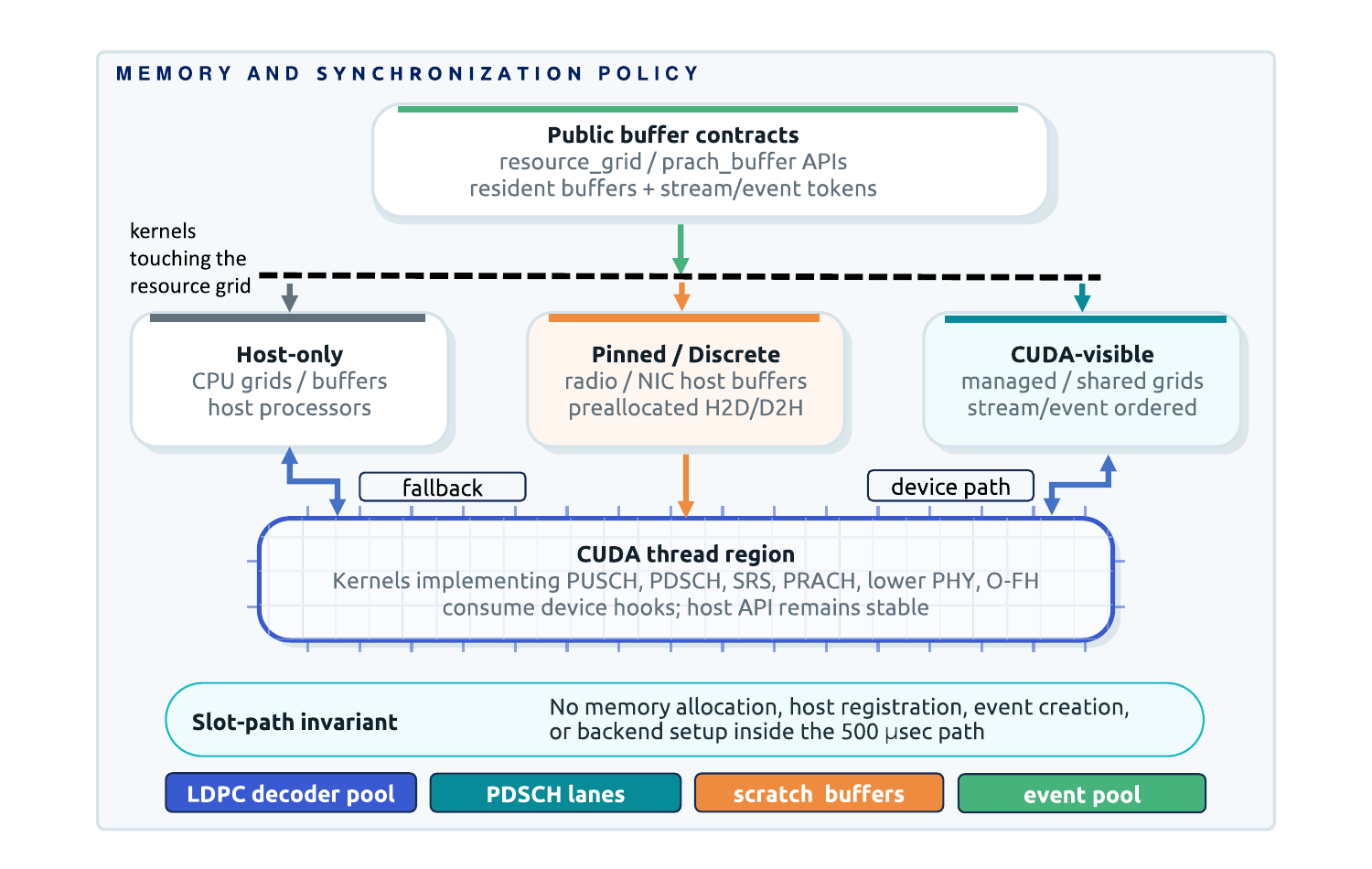}
\caption{Buffer policy abstraction. High-level code sees the same contracts while CUDA implementations choose CPU-only backing, pinned staging, or CUDA-visible shared/device backing.}
\label{fig:memory}
\end{figure*}

Three memory modes are supported (Fig.~\ref{fig:memory}). The three modes are a superset across platforms rather than a free choice on any one machine: which modes a given host supports is a property of that platform, chiefly whether a CUDA-capable GPU is present and whether it shares coherent memory with the host CPU, so the operative mode is dictated by the hardware available to the user. DeepSig's CUDA backend implements all three, so one codebase covers CPU-only hosts, discrete-GPU servers, and unified-memory platforms. In CPU-only mode, used when no compatible GPU is present or acceleration is disabled, all grids and PRACH buffers are host objects and the existing CPU processors run unchanged. In pinned/discrete mode, host grids remain authoritative and CUDA implementations use preallocated pinned staging buffers and explicit copies. This is the robust path for discrete GPUs or host-only radio packet buffers. In shared/unified mode, the grid or PRACH buffer is allocated with CUDA managed memory and advised for GPU access. This mode is preferred on integrated or coherent platforms such as the NVIDIA DGX Spark platform measured in this paper, because lower-PHY, O-FH, PUSCH, SRS, PRACH, and PDSCH can share the same allocation. Prefetching is left as an option because it can help tail latency in some downlink cases but hurt common uplink direct-managed timing.

Latency-critical allocation is moved out of the slot path. PUSCH decoders preallocate LDPC workspaces, rate-matcher buffers, scrambler sequence storage, resident softbit buffers, and output staging. O-FH CUDA compression handles retain device scratch buffers and pinned staging slots. The CUDA-visible grid owns a pool of producer-ready events so consecutive writers can record readiness without allocating events at runtime. Repeated \mode{cudaMalloc}, host registration, event creation, or managed-memory migration inside a 0.5 ms slot can dominate the kernels that follow.

\section{Accelerated Processing Blocks}
\subsection{PUSCH Receiver}
The accelerated PUSCH path, the PUSCH block in the uplink-receive region of Fig.~\ref{fig:pipeline}, is resident from grid input through LDPC and transport-block output. Eligibility requires SCH data, no UCI-on-PUSCH, and compatible demodulator and decoder implementations. When eligible, the CPU channel estimator path is bypassed; the CUDA E2E demodulator performs channel estimation, equalization, soft demapping, and metric accumulation. It returns a resident FP16 softbit pointer plus a CUDA completion event. The decoder stream waits on that event, then performs deinterleaving, rate recovery, LDPC decoding, CRC, and TB reassembly.

The LDPC integration includes several latency-specific choices. Decoder workspaces, rate-matcher state, and scrambling sequences are preallocated for worst-case dimensions. Uniform codeblock batches use GPU batch dematching. Resident paths fuse descrambling, deinterleaving, and dematching when possible and avoid LLR D2H/H2D transfers. The LDPC algorithm can be selected as \mode{boxplus}, \mode{min\_sum}, or \mode{auto}; auto uses a codeblock-count threshold so small batches can use the lower-latency box-plus path while larger half2 batches use min-sum where occupancy and memory behavior are better. Hard-decision, CB CRC, TB desegmentation, and TB CRC are performed on the GPU before a single payload/metadata synchronization.

The decoder also has a direct rate-match-to-LDPC path for large compatible batches. Rather than materializing every intermediate as an independent global-memory tensor, the rate-matched resident softbits can be consumed by an LDPC APP-initialization path that folds rate recovery into the decode preparation. When the transport block has nonuniform codeblock \(E\) or filler parameters, consecutive compatible runs are processed separately and decoded as a batch after their rate-recovered regions are populated. This preserves correctness for irregular TBs while keeping the common wideband cases on the fastest path.

\subsection{PDSCH Transmitter}
The PDSCH GPU block, at the head of the downlink-transmit path in Fig.~\ref{fig:pipeline}, accelerates the downlink transport-block path: LDPC encoding, rate matching, scrambling, modulation, layer mapping, precoding, and resource-grid mapping. The processor can use multiple lanes through \mode{pdsch\_acceleration\_nof\_lanes}; \mode{0} selects the default policy. A direct device-grid writer path writes mapped symbols into CUDA-visible DL grids. If a legacy grid is provided, a sidecar device grid can be materialized and synchronized back only where needed.

The downlink chain is deterministic once a PDU arrives, so there is no receiver sensitivity tradeoff and the latency question reduces to amortizing encoding and mapping work across codeblocks and layers. The CUDA implementation emphasizes codeblock batching, direct grid writes, and avoiding host-side symbol loops. For one-layer, small-bandwidth allocations, the GPU still pays fixed launch and synchronization cost; for four-layer and 100 MHz cases, the GPU path has enough parallelism to dominate.

\subsection{SRS, PRACH, and PUCCH}
The SRS estimator, shown beside PUSCH in the uplink-receive region of Fig.~\ref{fig:pipeline}, has a CUDA implementation using the same acceleration-mode policy. It can consume a CUDA-visible UL grid and emit channel-estimation diagnostics without changing the high-level SRS interfaces. PRACH acceleration exists in two places along the PRACH path of Fig.~\ref{fig:pipeline}: lower-PHY PRACH demodulation and upper-PHY PRACH detection. The PRACH buffer mirrors the resource-grid model with CUDA-visible managed storage and stream/event preparation hooks. PUCCH remains on the CPU path in the current implementation, which leaves a residency gap: when PUCCH is scheduled it can force host access to UL grid data that PUSCH and SRS would otherwise consume device-resident.

\subsection{Split-8 Lower PHY}
For split-8 SDR operation, the CUDA lower-PHY RX path (low-PHY RX in Fig.~\ref{fig:pipeline}) performs OFDM demodulation and writes directly into a CUDA-visible UL grid. The TX path performs OFDM modulation from the CUDA-visible DL grid. VkFFT is used for the transform-heavy CUDA lower-PHY, SRS, and PRACH blocks, avoiding a separate cuFFT dispatch model and keeping execution within \ocudu{}'s existing threading and stream orchestration. Acceleration can be independently disabled for TX, RX, and PRACH demodulation.

The split-8 path was designed for stable over-the-air operation with conventional USRP plugins. The radio still owns host sample buffers, but the GPU low-PHY RX path immediately converts those samples into the grid abstraction consumed by upper PHY. This keeps the later PUSCH and SRS stages resident even when the radio transport itself is not GPU-direct. On integrated GPUs, the RX path uses \emph{scattered mapped zero-copy}: each symbol$\times$port FFT window is host-registered as mapped memory and the load kernel reads host DRAM through per-batch device pointers, so multi-symbol batches skip both an intermediate host pack and an explicit H2D; tiny single-symbol cases keep a packed staging path under a size gate. On TX, the inverse path allows PDSCH output to remain on the accelerator until the final sample-buffer boundary.

\subsection{Split-7.2 O-FH Compression}
The O-FH acceleration layer, the O-FH RX and TX blocks at the resident-region boundaries of Fig.~\ref{fig:pipeline}, implements \mode{none} and BFP IQ compression/decompression for the common O-RAN bit widths. The CPU-auto baseline selects AVX on x86 or NEON on aarch64; on the platform measured in this paper, CPU-auto selects the ARM NEON implementations, not the scalar generic path. CUDA kernels include specializations for 9-bit and 12-bit packing, device-grid port batches, symbol batches, device-to-device compressed buffers, and host-buffer packet payloads. In RX, compressed packet payloads can be decompressed into a device-visible resource grid or PRACH buffer. In TX, a device-visible DL grid can be compressed directly into host-visible O-FH packet payload buffers or device buffers when an upstream transport path supports them.

The 9-bit BFP case is bandwidth efficient but difficult to accelerate: its payload is compact and bit-aligned awkwardly, so packing, unpacking, and memory layout dominate over arithmetic. The CUDA implementation uses warp-level PRB specializations and symbol/port batching to expose enough work per launch. Because these effects vary with shape, the benchmark matrix reports compression and decompression across bandwidths, ports, and bit widths.

\section{Validation Methodology}
The branch includes unit, integration, and benchmark coverage for the accelerated paths. Unit tests validate CUDA-visible resource-grid and PRACH-buffer interfaces, O-FH compression correctness, O-FH data-flow writers, lower-PHY PRACH demodulation, PRACH detection, and processor validators. Integration tests exercise PUSCH resident demodulation and decoding, PUSCH GPU/CPU comparison, PUSCH correctness and sensitivity sweeps, and PDSCH GPU E2E operation. Benchmark tools cover PUSCH, PDSCH, LDPC, PRACH, SRS, low-PHY demodulation, and O-FH compression; the PRACH benchmark can also emit latency and injected-sensitivity CSV files for plotting.

Correctness checks compare algorithmic output alongside timing. PUSCH latency tests decode the same generated transport blocks on CPU and GPU, compare pass/fail outcomes, check byte mismatches, and report CSI metric parity. Sensitivity sweeps use identical seeds and channel settings to estimate the BLER waterfall for CPU and GPU receivers. O-FH tests compare decompressed IQ against CPU references within quantization tolerance. Several optimizations, including FP16 softbits, box-plus/min-sum selection, fused rate dematching, and GPU hard decisions, could improve latency while quietly degrading link behavior, and the comparisons exist to catch that.

\section{Measurements}
\subsection{Setup}
Measurements were collected on an NVIDIA DGX Spark platform using an NVIDIA GB10 GPU and an aarch64 ARM host CPU containing Cortex-X925 and Cortex-A725 cores. Unless noted otherwise, binaries were built from the \ocudu{} CUDA-acceleration integration branch \mode{wg/nvcuda\_accel\_01} at commit \mode{3c89cdee}, which is the branch carrying the accelerated kernels evaluated here. Benchmark processes were pinned to the high-capacity CPU set 5--9 and 15--19, and the build used CUDA-enabled \ocudu{} benchmark binaries. PUSCH latency used the E2E pipeline benchmark with qam256 MCS 27, Type-1 double-symbol DM-RS, direct-managed RX grid mode, and pinned GPU output payload buffers. PDSCH latency used the GPU latency benchmark with qam256 MCS 27, double DM-RS, a sidecar CUDA device grid, and synchronized completion. Low-PHY RX results use 30 kHz SCS with 51, 106, and 273 PRB for 20, 40, and 100 MHz-class carriers; the table reports amortized per-symbol p50 latency from a 13-symbol same-CP batch with CUDA graphs enabled and scattered mapped zero-copy on the integrated GB10 path, which avoids synchronizing each OFDM symbol independently and skips host pack/H2D on multi-symbol windows. O-FH results are p50 values from the benchmark matrix with 14-symbol batches and synchronous \mode{device-device-symbol-batch} GPU mode. PRACH detector latency uses CUDA-visible PRACH buffers on the GPU path. The O-FH CPU baseline is the production \mode{cpu\_auto} CPU implementation, which resolves to NEON on this ARM platform.

Mode probes selected completed-work GPU paths for the tables: PUSCH direct-managed/pinned, PDSCH sidecar device grid with full codeblock batching, low-PHY batched VkFFT over same-CP symbols with scattered mapped zero-copy, synchronous O-FH device-to-device symbol batching, and PRACH CUDA-visible buffers; async O-FH enqueue timing is excluded.

All performance numbers in this paper were collected on \mode{wg/nvcuda\_accel\_01} (\mode{3c89cdee}), and the tables characterize latency of the benchmark binaries rather than establishing performance limits or full system throughput. OTA throughput also depends on scheduler policy, UE category and channel state, radio transport, fronthaul timing, HARQ behavior, and core-network path. What the measurements do show is that the GPU kernels and buffer paths are fast enough to move the L1 bottleneck for wideband and multilayer cases while maintaining CPU/GPU receiver parity in the measured PUSCH sensitivity sweep.

\subsection{Full-gNB Dummy-RU Scaling}
\ocudu{} also includes a built-in dummy radio unit for modem-level stress testing without RF hardware. The gNB still runs its DU high, scheduler, FAPI/DU-low, upper-PHY, and RU timing loops, while a synthetic RU supplies slot timing and request accounting for one connected UE per cell. This is an idealized resource-grid-boundary stress test: it excludes RF, split-8 OFDM, split-7.2 packet handling, and O-FH compression, but exercises the same factory-selected GPU PDSCH, PUSCH, SRS, PRACH, and CUDA-visible grid paths used by the live gNB.

Table~\ref{tab:dummyru} reports 100 MHz, 30 kHz SCS dummy-RU scaling for both 1T1R one-layer traffic and 4T4R traffic with the test UE RI set to four. The 4T4R sweep increases the downlink layer load while the synthetic uplink throughput in this topology remains at the one-layer scheduler point. Direct CUDA-visible managed UL/DL grids were used with PUSCH, PDSCH, SRS, and PRACH acceleration forced enabled; lower-PHY and O-FH acceleration were not part of this dummy-RU topology. Host runs used real-time privilege, a mixed big/LITTLE main pool on cores \(2\)--\(11\) and \(15\)--\(19\) (14 workers), PDSCH on the non-RT medium executor with max batch 1, \mode{OCUDU\_PDSCH\_CUDA\_GRAPHS=0}, a block-dependent PDSCH processor-pool override (24 for the lanes \(=1\) block, 64 for the lanes \(=3\) block), three-slot RU DL request margin, max processing delay 8, and extra DL resource-grid depth 48. Both campaigns enable test-mode auto-ACK delay~5 as a multi-cell HARQ stabilizer (MAC synthesizes delayed ACK/CRC; L1 still processes full-buffer PDSCH/PUSCH). The 4T4R campaign sets PDSCH concurrency and acceleration lanes equal to the sector count. The 1T1R campaign uses a \emph{scale-dependent} PDSCH lane policy (forcing lanes \(=N\) reintroduces NOK/min-rate fails on one-layer traffic): lanes \(=1\) for one through four sectors, then a fixed modest concurrency of three lanes for five and higher sectors. Three lanes overlap multi-cell encodes without the lanes \(=N\) HARQ thrash and, relative to lanes \(=1\) at the same \(N\), lower measured CPU/GPU utilization by finishing DL work with less serial queueing. The 1T1R path also sets \mode{OCUDU\_LDPC\_BOXPLUS=0}. Scheduler throughput is averaged over the final steady rows; RU-late counters are deltas over the RU measurement window.

\begin{table*}[t]
\caption{Full-gNB dummy-RU GPU sector scaling, 100 MHz, on \mode{wg/nvcuda\_accel\_01} (\mode{3c89cdee}). 1T1R rows 1--4 use one PDSCH lane and a PDSCH processor pool of 24; rows 5 and above use three fixed lanes (not lanes \(=N\)) and a pool of 64; 4T4R uses lanes \(=N\). CPU/GPU columns are host busy and device util over the RU telemetry window, comparable within a recipe block rather than a kernels-only load model. The daggered row failed the campaign's own validation because some cells never served traffic; its DL/cell is the mean over all cells including those. The ten- and eleven-sector points at pool~64 did not validate on any attempt; the \(N=12\) row is a single validated observation (see text).}
\label{tab:dummyru}
\centering
\scriptsize
\setlength{\tabcolsep}{3pt}
\resizebox{\textwidth}{!}{%
\begin{tabular}{@{}lrrrrrrrrrr@{}}
\toprule
Config & Sectors & DL/cell & UL/cell & Agg. DL & Agg. UL & Sched. NOK & RU late DL/UL/PRACH & CPU avg/max & GPU avg/max & RSS \\
 & & (Mb/s) & (Mb/s) & (Mb/s) & (Mb/s) & avg/max & & & & \\
\midrule
1T1R & 1 & 404.0 & 102.0 & 404.0 & 102.0 & 0.00/0.00\% & 0.000/0.015/0.000\% & 21.3/25.6\% & 10.4/12.0\% & 5.6 GB \\
1T1R & 2 & 404.0 & 102.0 & 808.0 & 204.0 & 0.00/0.00\% & 0.000/0.015/0.059\% & 26.1/28.0\% & 20.2/26.0\% & 6.0 GB \\
1T1R & 3 & 404.0 & 102.0 & 1212.0 & 306.0 & 0.00/0.00\% & 0.007/0.015/0.000\% & 30.9/35.7\% & 30.6/39.0\% & 6.5 GB \\
1T1R & 4 & 404.0 & 102.0 & 1616.0 & 408.0 & 0.00/0.00\% & 0.003/0.000/0.000\% & 35.6/42.5\% & 39.8/52.0\% & 7.0 GB \\
1T1R & 5 & 404.0 & 102.0 & 2020.0 & 510.0 & 0.00/0.00\% & 0.005/0.021/0.024\% & 39.5/49.1\% & 33.5/45.0\% & 8.3 GB \\
1T1R & 6 & 404.0 & 102.0 & 2424.0 & 612.0 & 0.00/0.00\% & 0.002/0.012/0.000\% & 44.4/54.1\% & 40.6/55.0\% & 8.9 GB \\
1T1R & 7 & 403.7 & 102.0 & 2826.0 & 714.0 & 0.06/0.00\% & 0.002/0.028/0.032\% & 47.3/57.2\% & 44.8/60.0\% & 9.5 GB \\
1T1R & 8 & 403.6 & 102.0 & 3228.8 & 815.8 & 0.06/0.00\% & 0.041/0.028/0.007\% & 48.3/58.3\% & 47.6/64.0\% & 10.1 GB \\
1T1R & 9 & 403.3 & 102.0 & 3630.0 & 917.8 & 0.11/0.00\% & 0.256/0.029/0.007\% & 49.3/60.4\% & 51.0/69.0\% & 10.7 GB \\
1T1R\(\dagger\) & 10 & 282.7 & 71.4 & 2826.8 & 714.0 & 0.04/0.00\% & 0.020/0.057/0.006\% & 46.2/56.7\% & 44.5/61.0\% & 11.3 GB \\
1T1R & 12 & 403.3 & 101.9 & 4839.6 & 1222.8 & 0.12/0.00\% & 3.423/0.031/0.005\% & 49.6/62.9\% & 58.0/83.0\% & 12.6 GB \\
4T4R & 1 & 1400.0 & 102.0 & 1400.0 & 102.0 & 0.00/0.00\% & 0.000/0.000/0.000\% & 23.2/27.2\% & 5.0/6.0\% & 6.2 GB \\
4T4R & 2 & 1400.0 & 102.0 & 2800.0 & 204.0 & 0.01/0.00\% & 0.000/0.000/0.000\% & 28.9/32.7\% & 9.2/12.0\% & 7.3 GB \\
4T4R & 3 & 1400.0 & 102.0 & 4200.0 & 306.0 & 0.00/0.00\% & 0.000/0.000/0.000\% & 34.6/39.0\% & 12.6/24.0\% & 8.6 GB \\
4T4R & 4 & 1400.0 & 102.0 & 5600.0 & 408.0 & 0.00/0.00\% & 0.004/0.000/0.000\% & 39.6/47.3\% & 15.5/27.0\% & 10.0 GB \\
4T4R & 5 & 1400.0 & 102.0 & 7000.0 & 510.0 & 0.00/0.00\% & 0.000/0.015/0.000\% & 45.4/58.8\% & 19.1/28.0\% & 11.5 GB \\
4T4R & 6 & 1400.0 & 102.0 & 8400.0 & 612.0 & 0.00/0.00\% & 0.002/0.000/0.000\% & 49.6/62.5\% & 20.9/30.0\% & 13.2 GB \\
\bottomrule
\end{tabular}
}
\vspace{-1mm}
\end{table*}

Multi-sector scaling in both topologies is governed by host-side slot orchestration and PDSCH admission policy rather than by kernel throughput. The \textbf{4T4R} campaign uses one PDSCH concurrency lane \emph{per sector} (lanes \(=N\)) and test-mode auto-ACK delay~5 to suppress multi-cell HARQ startup storms. It holds the MCS-27 buffer ceiling of 1400~Mb/s DL and 102~Mb/s UL on every cell through \textbf{six sectors} (8.4~Gb/s aggregate DL) with RU DL-late at or below 0.004\% on every row and device utilization peaking at only 30\%. The 50\% host CPU average at six sectors---against a 21\% GPU average---shows that this topology is bounded by host-side slot orchestration rather than by the accelerator.

The \textbf{1T1R} campaign is admission-limited rather than GPU-FLOP-limited, and two distinct configuration limits set its operating envelope. Forcing lanes \(=N\) produces scheduler NOK and near-miss min-rate failures on one-layer traffic, so the table never uses per-sector lane counts: lanes \(=1\) serves one through four sectors and a fixed modest concurrency of \textbf{three} PDSCH lanes serves five and above. The second limit is PDSCH processor-pool depth, which must scale with the product of lanes and cells: a pool of 24 is sufficient for the lanes \(=1\) block but not once three lanes are active across nine or more cells, where the ninth cell is never issued traffic. At fixed \(N=9\) the dependence is reproducible in both directions --- a pool of 24 fails on five independent attempts while a pool of 48 serves the ninth cell immediately (\(3634\)~Mb/s aggregate, RU DL-late 0.176\%). With the pool at 64 the recipe holds full-rate 404/102 through \textbf{nine sectors} (3.63~Gb/s aggregate DL) with RU DL-late at or below 0.256\%, and host CPU and GPU utilization rise monotonically inside the lanes \(=3\) block to roughly 49\% and 69\% respectively at nine sectors.

Beyond nine sectors a different failure dominates, and it is not a capacity limit. Sweeping \(N=5\ldots12\) at pool~64 gives a non-monotonic outcome: ten and eleven sectors fail while \textbf{twelve} validates at 4.84~Gb/s aggregate DL with 3.42\% RU DL-late and 83\% peak device utilization. No pool-sizing or throughput model can produce that ordering, since twelve sectors demand strictly more of both than ten. The signature instead points to an intermittent multi-cell startup fault: on failing runs a \emph{varying} subset of cells reports exactly 0.0~Mb/s while RU DL-late on the same run stays at or near 0.000\%, so those cells are never issued downlink work rather than missing deadlines. Four consecutive ten-sector attempts at pool~64 failed with seven, three, four, and then a differing subset of cells served. We therefore report nine full-rate 1T1R sectors as the highest \emph{reliably reproducible} operating point and treat the twelve-sector row as a single validated observation that establishes headroom without establishing reliability. This failure mode also constrains how any row in Table~\ref{tab:dummyru} should be read: because the campaign runner retries a point until it validates, a fault this intermittent makes a clean row partly a selection over attempts rather than evidence of a steady state, so the comparison rests on per-point attempt counts rather than on best-of results. Auto-ACK is used only as a MAC test-mode HARQ stabilizer; it does not bypass PDSCH/PUSCH GPU processing, and NOK\(\approx 0\) under that mode should not be read as a realistic feedback metric.

\begin{table}[t]
\caption{PUSCH E2E Latency and Correctness on DGX Spark, qam256 MCS 27. Pass counts are decoded frames out of 200 measured iterations.}
\label{tab:pusch}
\centering
\footnotesize
\begin{tabularx}{\columnwidth}{@{}lrrrrr@{}}
\toprule
Case & TBS kb & CPU \us & GPU \us & Spd. & Pass CPU/GPU \\
\midrule
20 MHz 1L & 54.3 & 334.0 & 183.9 & 1.82x & 200/200 \\
20 MHz 4L & 217.1 & 1451.6 & 253.5 & 5.73x & 182/182 \\
40 MHz 1L & 112.6 & 484.3 & 181.8 & 2.66x & 200/200 \\
40 MHz 4L & 451.0 & 3046.7 & 296.1 & 10.29x & 181/181 \\
100 MHz 1L & 295.2 & 980.3 & 245.2 & 4.00x & 200/200 \\
100 MHz 4L & 1179.9 & 4459.8 & 586.8 & 7.60x & 179/180 \\
\bottomrule
\end{tabularx}
\vspace{-1mm}
\end{table}

Table~\ref{tab:pusch} reports mean PUSCH total latency over 200 measured iterations on \mode{wg/nvcuda\_accel\_01} at \mode{3c89cdee} (live modem off). Byte comparisons showed zero mismatches in every case. Five of the six rows agree exactly on decoded-frame counts; the 100\,MHz four-layer row differs by a single frame out of 200 (GPU 180, CPU 179) at an operating point whose short-run frame-error rate is already near 10\%, which is finite-BLER variation between two independent decoders rather than a decode error. CSI metrics were populated on the GPU path, with measured SINR deltas no larger than 0.02 dB. Peak PUSCH speedup is \(10.29\times\) at 40\,MHz 4L; the 100\,MHz 4L GPU mean remains \(\sim\)587\,\us{} with LDPC still dominant.

Stage instrumentation on the same runs locates that cost. For the 100\,MHz four-layer path, LDPC decode alone accounts for roughly two thirds of the GPU mean (377.6\,\us{} of 586.8\,\us{}), the host-side decode and demodulation wrappers contribute most of the remainder (123.1 and 48.3\,\us{}), CRC checking adds 14.6\,\us{}, and the device-to-host transfer is only 5.2\,\us{} once the path is resident. That distribution sets the priority for further PUSCH work: LDPC kernel quality and decode-path fusion dominate absolute microseconds, whereas small one-layer grants sit closer to a fixed launch and setup floor.

\begin{table}[t]
\caption{PDSCH Latency on DGX Spark, qam256 MCS 27}
\label{tab:pdsch}
\centering
\footnotesize
\begin{tabularx}{\columnwidth}{@{}lrrrr@{}}
\toprule
Case & TBS kb & CPU p50 \us & GPU p50 \us & Spd. \\
\midrule
20 MHz 1L & 54.3 & 28.8 & 57.1 & 0.50x \\
20 MHz 4L & 217.1 & 112.9 & 69.4 & 1.63x \\
40 MHz 1L & 112.6 & 58.3 & 66.4 & 0.88x \\
40 MHz 4L & 451.0 & 230.4 & 91.2 & 2.53x \\
100 MHz 1L & 295.2 & 148.2 & 89.9 & 1.65x \\
100 MHz 4L & 1179.9 & 603.7 & 226.1 & 2.67x \\
\bottomrule
\end{tabularx}
\vspace{-1mm}
\end{table}

PDSCH speedups increase with TB size and layer count because LDPC encoding, rate matching, and mapping expose larger batches and amortize kernel launch overhead. With resident TB encoder lanes and CUDA graphs disabled, the 20 MHz and 40 MHz one-layer rows remain at or below CPU p50 (0.50$\times$ and 0.88$\times$); four-layer rows and 100 MHz rows expose enough batching for the GPU path to win (up to \(2.67\times\) at 100 MHz 4L in this sidecar device-grid mode). For deployments that prioritize end-to-end residency, the narrow one-layer cases can still stay on the GPU, but their standalone latency is dominated by fixed launch, CRC, LDPC setup, and completion overhead.
Multi-cell dummy-RU scaling (Table~\ref{tab:dummyru}) further shows that PDSCH concurrency/lanes and CUDA graph policy affect \emph{system} late counters even when per-PDU GPU encode is not FLOP-bound.

\begin{table}[t]
\caption{DGX Spark Split-8 Low-PHY RX OFDM Demodulation. Left: 1-symbol completed-work p50. Right: 13-symbol same-CP batch amortized per symbol. The GPU floor is visible on the left and amortized away on the right.}
\label{tab:lowphy}
\centering
\scriptsize
\setlength{\tabcolsep}{3.5pt}
\begin{tabular}{@{}lrrrrrrr@{}}
\toprule
& & \multicolumn{3}{c}{1-symbol completed work} & \multicolumn{3}{c}{13-symbol amortized} \\
\cmidrule(lr){3-5}\cmidrule(lr){6-8}
Carrier & Ports & CPU \us & GPU \us & Spd. & CPU \us & GPU \us & Spd. \\
\midrule
20 MHz & 1 & 1.6 & 8.1 & 0.20x & 1.6 & 0.9 & 1.73x \\
20 MHz & 2 & 3.3 & 10.8 & 0.31x & 3.2 & 0.9 & 3.41x \\
20 MHz & 4 & 6.6 & 13.8 & 0.48x & 6.5 & 1.0 & 6.33x \\
40 MHz & 1 & 3.4 & 8.6 & 0.40x & 3.4 & 0.9 & 3.59x \\
40 MHz & 2 & 6.9 & 11.3 & 0.61x & 6.7 & 1.0 & 6.59x \\
40 MHz & 4 & 13.7 & 12.3 & 1.11x & 13.5 & 1.1 & 12.28x \\
100 MHz & 1 & 7.5 & 9.8 & 0.77x & 7.2 & 0.9 & 7.62x \\
100 MHz & 2 & 14.8 & 12.3 & 1.20x & 14.6 & 1.2 & 11.81x \\
100 MHz & 4 & 29.5 & 12.3 & 2.40x & 29.4 & 1.5 & 19.72x \\
\bottomrule
\end{tabular}
\vspace{-1mm}
\end{table}

Table~\ref{tab:lowphy} shows the effect of slot-shaped batching on split-8 low-PHY acceleration. CPU FFTW is very fast for a single one-port symbol, and a GPU path pays a completed-work launch/synchronization floor that an isolated symbol microbenchmark never recovers. Batching 13 same-CP OFDM symbols into one batched VkFFT dispatch and one synchronization amortizes that floor, makes every measured carrier/port row faster, and produces a CUDA-visible grid for PUSCH and SRS.
Scattered mapped zero-copy on the integrated GB10 path, which skips both the multi-window host pack and an explicit H2D, is the main reason the 100\,MHz 4-port amortized row stays near \(1.5\,\us{}/\)symbol (\(\sim\)19.7$\times$ vs CPU).

Because the two operating points answer different questions, Table~\ref{tab:lowphy} reports both from the same binary rather than the amortized figure alone. On 20\,MHz one-port the ranking inverts: 1.6 versus 8.1\,\us{} for a single symbol (0.20$\times$), but 1.6 versus 0.9\,\us{} per symbol once batched (1.73$\times$). The GPU per-symbol cost stays nearly flat across every carrier and port count while the CPU cost grows with both, so the crossover is set by how much work is batched per launch rather than by the carrier. Dummy-RU multi-cell campaigns do not instantiate split-8 lower-PHY, so Table~\ref{tab:dummyru} must not be read as a split-8 system latency result.

\begin{table}[t]
\caption{DGX Spark O-FH BFP Compression/Decompression, CPU-auto NEON vs CUDA}
\label{tab:ofh}
\centering
\footnotesize
\begin{tabularx}{\columnwidth}{@{}llrrrr@{}}
\toprule
Case & Path & Bits & CPU \us & GPU \us & Spd. \\
\midrule
20M 4p & TX & 9 & 14.7 & 7.4 & 1.99x \\
20M 4p & RX & 9 & 13.2 & 7.8 & 1.69x \\
100M 4p & TX & 9 & 80.0 & 14.0 & 5.71x \\
100M 4p & RX & 9 & 67.2 & 8.6 & 7.81x \\
20M 4p & TX & 12 & 162.6 & 7.0 & 23.23x \\
20M 4p & RX & 12 & 134.6 & 6.0 & 22.43x \\
100M 4p & TX & 12 & 884.5 & 12.3 & 71.91x \\
100M 4p & RX & 12 & 731.4 & 8.0 & 91.42x \\
\bottomrule
\end{tabularx}
\vspace{-1mm}
\end{table}

In Table~\ref{tab:ofh}, event-based completion and shape-specific BFP9/BFP12 kernel selection move every measured O-FH row above parity. BFP9 remains more bit-manipulation intensive, and the narrow 20 MHz rows have the smallest standalone margins (1.7--2.0$\times$); BFP12 and 100 MHz rows amortize fixed costs strongly (up to 91.4$\times$ RX decompress for BFP12) and preserve end-to-end residency.
Across the matrix the GPU stays roughly flat in a 6--14\,\us{} band while the CPU cost scales with both bandwidth and bit width, which is what turns a modest 20\,MHz advantage into a large 100\,MHz one.

\begin{table}[t]
\caption{PRACH detector latency and injected-sweep sensitivity}
\label{tab:prach}
\centering
\scriptsize
\setlength{\tabcolsep}{2.7pt}
\begin{tabular}{llrrrrr}
\toprule
Fmt. & Ports & Pre. & CPU p50 & GPU p50 & Spd. & \(\ge90\%\) SNR \\
     &       &      & (\us)   & (\us)   &      & CPU/GPU \\
\midrule
B4 & 1 & 64 & 64.4 & 12.5 & 5.15x & -14/-14 dB \\
B4 & 4 & 64 & 175.3 & 15.5 & 11.31x & -22/-22 dB \\
0  & 1 & 64 & 276.9 & 14.0 & 19.78x & -14/-14 dB \\
0  & 4 & 64 & 576.5 & 20.0 & 28.82x & -20/-20 dB \\
\bottomrule
\end{tabular}
\end{table}

Table~\ref{tab:prach} reports upper-PHY PRACH detector measurements using CUDA-visible PRACH buffers on the GPU path, with true-IDFT correlation, CUDA graphs, and adaptive mapped/device result staging. The sensitivity column is retained from the prior injected frequency-domain detector sweep (CPU/GPU thresholds matched). Light short-PRACH rows land near $\sim$12--16\,\us{}; heavy format-0 multiport finishes near $\sim$20\,\us{}, scaling to 28.8$\times$ vs CPU. Keeping the detector on the GPU also lets split-8 lower-PHY PRACH demodulation and split-7.2 O-FH decompression hand it a CUDA-visible PRACH buffer without staging the PRACH window through a host-only copy.

\begin{figure*}[t]
\centering
\includegraphics[width=\textwidth]{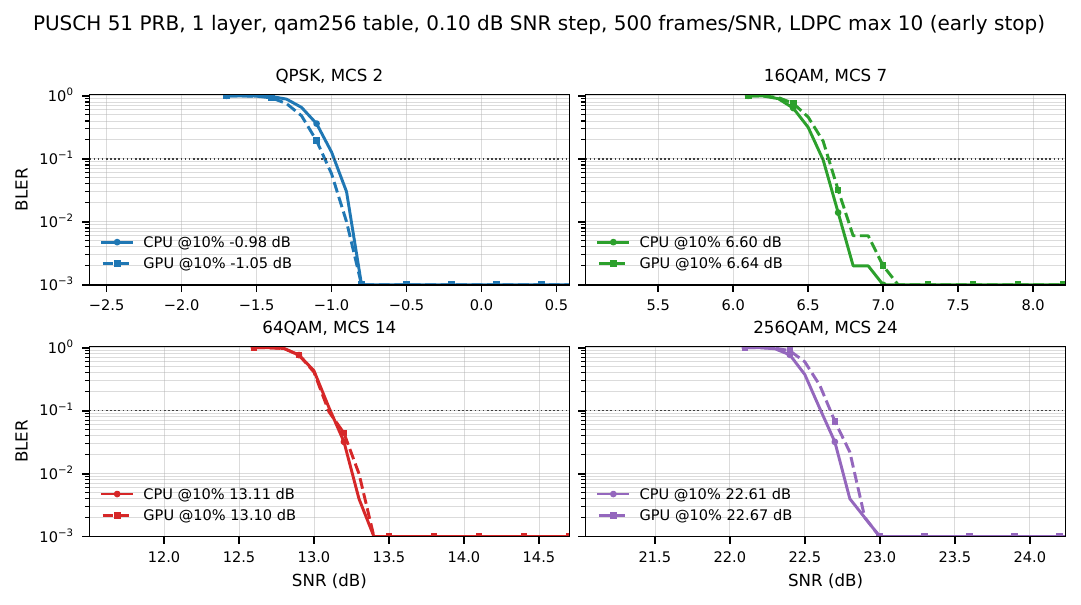}
\caption{PUSCH BLER sweeps for one representative MCS per modulation on a 51 PRB, one-layer qam256-table configuration, measured on \mode{wg/nvcuda\_accel\_01} (\mode{3c89cdee}). CPU and GPU decode identical frames with common per-frame noise realizations replayed at every SNR point, using a coarse locating scan followed by a 0.10~dB fine sweep at 500 frames per point and LDPC max-10 early stopping. Measured CPU/GPU 10\% BLER thresholds are $-0.983/-1.045$ (QPSK), $6.597/6.636$ (16QAM), $13.109/13.097$ (64QAM), and $22.606/22.671$~dB (256QAM), agreeing to within 0.064~dB on every case.}
\label{fig:sensitivity}
\end{figure*}

The PUSCH BLER sweeps in Fig.~\ref{fig:sensitivity} use identical seeds, channel settings, and LDPC iteration limits for CPU and GPU, and replay common per-frame noise realizations at every SNR point so the CPU/GPU comparison is paired rather than independently sampled. The CPU/GPU 10\% BLER thresholds match within 0.064~dB across the QPSK, 16QAM, 64QAM, and 256QAM representative cases, indicating that the LDPC, softbit quantization, demapping, and rate-recovery optimizations preserve receiver quality.

\section{Operational Integration}
Operational profiles set acceleration modes in YAML rather than application CUDA code. Split-8 profiles can set lower-PHY TX/RX, PRACH demodulation, PUSCH, PDSCH, SRS, and PRACH detection to \mode{auto}; split-7.2 profiles can do the same for O-FH compression. CUDA builds select GPU factories, while CPU-only builds fall back unless a mode is explicitly \mode{enabled}.

The CUDA-accelerated implementation is available as open-source software in the \ocudu{} hardware-acceleration working-group repository, where the results reported here correspond to the \mode{wg/nvcuda\_accel\_01} branch, and is being upstreamed into the main \ocudu{} project~\cite{ocudu-wg1,ocudu-gitlab}. Because the acceleration backend is in-tree, the CPU and GPU implementations share configuration, tests, build options, and integration paths, which reduces the risk that accelerator support diverges from the deployable modem.

Deployment tuning targets slot-path latency and jitter across the chain rather than individual micro-benchmarks. Narrow one-layer downlink grants, small BFP9 packets, and one-port FFT cases sit near the launch-overhead floor, yet keeping them on the GPU preserves residency and avoids larger copies before PDSCH, O-FH, PUSCH, or SRS. Policy selection is platform-aware: CPU-only without CUDA, pinned staging at unavoidable host radio boundaries, and CUDA-visible shared allocations where coherent GPU access is sustainable.

\section{Discussion}
Across the measurements, the GPU path performs best when it avoids host round trips and processes enough work per launch: resident demodulation and LDPC batching for PUSCH, larger transport blocks and direct grid mapping for PDSCH, and multi-port, multi-symbol batches for O-FH. The split is consistent across the suite. High-throughput PUSCH, wide PRACH, and BFP12 O-FH sit far above parity, while narrow one-layer PDSCH, BFP9 at 20\,MHz, and single-symbol low-PHY sit near or below a floor set by launch overhead and bit-packing efficiency rather than by arithmetic throughput.

Where absolute latency remains is specific to each block, and the stage data locate it rather than infer it. For PUSCH, LDPC decode is 64.3\% of the 100\,MHz four-layer GPU mean and 53.2\% at one layer, so further gains depend on LDPC kernel quality and decode-path fusion far more than on demodulation micro-optimizations; the 20\,MHz one-layer case reaches only 1.82$\times$ precisely because its demodulation wrapper and fixed setup are comparable in size to its LDPC work. PDSCH shows the mirror image: narrow one-layer rows still lose to NEON on standalone latency (0.50--0.88$\times$) while multi-layer and wide rows reach 2.67$\times$, and the multi-cell dummy-RU results show that system late counters are usually admission- and lane-limited rather than encode-limited, since GPU utilization can sit well below saturation while RU DL-late rises. For split-8, amortizing a 13-symbol batch reaches 19.7$\times$ on 100\,MHz four-port after scattered mapped zero-copy, yet the 1-symbol completed-work columns of the same table still show an 8--14\,\us{} GPU floor that inverts the ranking on narrow carriers; batching a full slot, or accepting device-resident IQ directly from the radio, is the remaining architectural lever rather than a further kernel optimization. O-FH is largely solved for BFP12 (22--91$\times$) and remains an optimization target for the more bit-manipulation-intensive BFP9 path (1.7--7.8$\times$). The PRACH detector is strong on format-0 multi-port (28.8$\times$) and reaches 5.2$\times$ on format-B4 single-port, where the fixed detector cost dominates a small correlation.

A methodological point follows from these numbers. The dummy-RU campaign exercises the full gNB with GPU PDSCH, PUSCH, SRS, and PRACH over managed grids, but it does not instantiate split-8 OFDM or O-FH factories, so it cannot be read as a split-8 or fronthaul system result. The microbench tables remain the right place to quote low-PHY and O-FH kernel latency, and dummy-RU the right place to quote multi-cell throughput and late rates under concurrent uplink and downlink load. Equally, an amortized per-symbol figure and a 1-symbol completed-work figure answer different questions, and reporting only the former would overstate what the accelerator does for a narrow carrier.

The current design still has known gaps. PUCCH remains CPU-only. NIC-to-GPU packet ingress and egress are not fully device-resident today, so split-7.2 packet headers and host Ethernet buffers remain a boundary. Some downlink packetization paths still need host-visible packet payloads. The buffer interfaces already distinguish CPU-only, pinned/discrete, and CUDA-visible managed modes, however, so future NIC-direct or GPUDirect-style integration can extend these boundaries without rewriting upper PHY.

For AI-RAN, the main contribution of this work is tensor locality. GPU-accelerated AI-native receivers, neural channel estimators, learned link adaptation features, SRS/DM-RS sensing pipelines, and other physical-layer AI functions need direct access to grids, softbits, channel estimates, and decoded metadata, and a GPU-resident \ocudu{} PHY lets these models run beside the standards-compliant kernels on the same accelerator with stream/event dependencies in place of file, socket, or host-memory side channels. This supports AI-native air-interface work in 5G Advanced and 6G, where learned processing belongs inside the L1 pipeline~\cite{hoydis2021ainative}.

The same abstraction keeps the software maintainable: public high-level files do not need to know whether a grid is backed by scalar CPU memory, pinned host memory, managed memory, or a device allocation. They ask whether a device view is supported and otherwise use the existing host path. Future acceleration candidates such as PUCCH, CSI report processing, deeper fronthaul packet batching, and NIC-direct packet buffers can follow the same pattern.

\section{Conclusion}
The CUDA-accelerated \ocudu{} stack now covers the major latency-critical L1 paths: PDSCH, PUSCH, SRS, PRACH demodulation/detection, split-8 lower PHY, and split-7.2 O-FH compression. The implementation keeps the CUDA-based backend behind factories and specialized component implementations, while resource-grid and PRACH-buffer extensions provide a small, reusable mechanism for device residency. On the measured DGX Spark GB10/ARM platform, representative results show up to 10.3$\times$ PUSCH, 2.7$\times$ PDSCH, 19.7$\times$ amortized split-8 low-PHY (scattered mapped zero-copy), 91.4$\times$ O-FH BFP12, and 28.8$\times$ PRACH detector speedups, with PUSCH sensitivity preserved in the tested sweep and multi-cell dummy-RU scale through nine full-rate 1T1R sectors (3.63~Gb/s aggregate DL) and six full-rate 4T4R sectors (8.4~Gb/s aggregate DL). Nine-sector 1T1R operation requires a PDSCH processor pool sized to the lane and cell counts, and a single twelve-sector run validated at 4.84~Gb/s, so the ceiling above nine sectors is set by an intermittent multi-cell startup fault---cells that are never issued downlink work, with RU-late counters near zero---rather than by kernel throughput or admission capacity. Making that startup path deterministic is the prerequisite for reporting a scaling curve past nine sectors. The remaining work is to cut PUSCH LDPC and narrow-PDSCH fixed cost, close PUCCH and deeper fronthaul packet residency, and expose GPU-resident tensors to AI-RAN applications without compromising real-time L1 scheduling.

\appendices
\section{Measurement Configuration}
The dummy-RU measurements in Table~\ref{tab:dummyru} used managed CUDA-visible UL/DL grids with PUSCH, PDSCH, SRS, and PRACH acceleration forced enabled; split-8 lower-PHY and split-7.2 O-FH factories were not instantiated in this synthetic-RU topology. Hosts ran \mode{performance} governors with the main pool on cores \(2\)--\(11\) and \(15\)--\(19\) (14 workers), PDSCH on the non-RT medium executor with max batch~1, CUDA graphs disabled for PDSCH, a PDSCH processor-pool size of~24 for the lanes \(=1\) rows and~64 for the lanes \(=3\) rows, a three-slot RU DL request margin, max processing delay~8, extra DL resource-grid depth~48, and test-mode auto-ACK delay~5 as a multi-cell HARQ stabilizer. The selected 1T1R rows use ten PUSCH/SRS workers with one PDSCH lane for one through four sectors and three fixed lanes for five sectors and above, never lanes \(=N\); the 4T4R rows use PDSCH concurrency equal to the sector count. Telemetry sampled scheduler, RU, CPU/RSS, and GPU metrics every 500~ms, and the daggered row should be read as a startup-fault diagnostic rather than a clean operating point or a saturation point. The microbenches in Tables~\ref{tab:pusch}--\ref{tab:prach} were pinned to cores 5--9 and 15--19 on a CUDA Release build: PUSCH over 200 iterations of completed end-to-end work, PDSCH over 200 iterations across five runs reported as the median run p50, low-PHY over 3000 repetitions for the 1-symbol and 13-symbol same-CP batch points with graphs on and scattered mapped zero-copy on the integrated GPU, O-FH as 14-symbol device-to-device symbol batches with synchronous completion, and PRACH with CUDA-visible buffers and graphs on. The O-FH CPU baseline is the production \mode{cpu\_auto} implementation, which resolves to NEON on this platform. The sensitivity sweep of Fig.~\ref{fig:sensitivity} used a coarse locating scan followed by a 0.10~dB fine sweep at 500 frames per SNR point, with CPU and GPU receivers driven from the same generated frames and noise realizations. Microbench latency runs and dummy-RU campaigns were never executed concurrently: the gNB main pool and the benchmark CPU set overlap, and sharing them produces dead cells and inflated RU-late counters that mimic a scaling limit.

\section*{Acknowledgment}
This work was supported in part by the U.S. DoD OUSD(R\&E) FutureG Office through the National Spectrum Consortium (NSC) Spectrum Forward OTA, including support for contributions to the Linux Foundation's \ocudu{} open-source RAN project.

\bibliographystyle{IEEEtran}
\bibliography{references}

\end{document}